\documentclass{aa}  

\usepackage{graphicx}
\usepackage{txfonts}
\usepackage[]{natbib}  
\usepackage{array}
\usepackage{colortbl}
\usepackage{multirow}
\usepackage{gensymb}
\usepackage{breqn}
\usepackage{fancyvrb}

\usepackage{booktabs}
\usepackage{amsmath}
\usepackage{siunitx}
\usepackage{subfig}
\newcommand{\iu}{{i\mkern1mu}}

\begin{document}

\title{Probing the internal magnetism of stars\\ using asymptotic magneto-asteroseismology}


   \author{S. Mathis$^{ *}$ 
          \inst{1}
          \and
          L. Bugnet$^{ *}$ \inst{1}
          \and
          V. Prat \inst{1}
          \and
          K. Augustson \inst{1}
          \and
          S. Mathur \inst{2,3}
          \and
          R. A. Garcia \inst{1}
          }

\institute{
D\'epartement d'Astrophysique-AIM, CEA/DRF/IRFU, CNRS/INSU, Universit\'e Paris-Saclay, Universit\'e Paris-Diderot, Universit\'e de Paris, F-91191 Gif-sur-Yvette, France\\
*: S. Mathis and L. Bugnet equally contributed to this work; 
\email{stephane.mathis@cea.fr;lisa.bugnet@cea.fr}
\and
Instituto de Astrof\'{\i}sica de Canarias, E-38200, La Laguna, Tenerife, Spain
\and 
Universidad de La Laguna, Dpto. de Astrof\'{\i}sica, E-38205, La Laguna, Tenerife, Spain}

   \date{Received/accepted}

 
  \abstract
   {Our knowledge of the dynamics of stars has undergone a revolution thanks to the simultaneous large amount of high-quality photometric observations collected by space-based asteroseismology and ground-based high-precision spectropolarimetry. They allowed us to probe the internal rotation of stars and their surface magnetism in the whole Hertzsprung-Russell diagram. However, new methods should still be developed to probe the deep magnetic fields in those stars.}
   {Our goal is to provide seismic diagnoses that allow us to sound the internal magnetism of stars.}
   {Here, we focus on asymptotic low-frequency gravity modes and high-frequency acoustic modes. Using a first-order perturbative theory, we derive magnetic splittings of their frequencies as explicit functions of stellar parameters.}
   {As in the case of rotation, we show how asymptotic gravity and acoustic modes can allow us to probe the different components of the magnetic field in the cavities where they propagate. This demonstrates again the high potential of using mixed-modes when this is possible.}
   {}

   
   \authorrunning{S. Mathis et al.}  
             
   \maketitle
%
\section{Introduction}

Along with rotation, magnetism is one of the two corner stones of stellar dynamics. For every type of stars, it deeply modifies their interactions with their surrounding environment and their evolution, in particular through the transport of angular momentum and chemicals they potentially trigger in their interiors \citep[e.g.][]{BrunBrowning2017,SpruitBraithwaite2017}.
In this framework, two revolutions have occurred in stellar physics over the last two decades. 

The first one is space-based helio- and asteroseismology \citep[e.g.][]{JCD2015,GarciaBallot2019,Aerts2019}. They have allowed us to probe with high precision the structure, the internal rotation, and the magnetic activity of the Sun and of stars. Helioseismology analysis done with spaced-based instruments (i.e. the MDI \citep[Michelson Doppler Imager;][]{Scherreretal1995} and GOLF \citep[Global Oscillations at Low Frequencies;][]{Gabrieletal1995} instruments onboard the SOHO \citep[SOlar and Heliospheric Observatory;][]{Domingoetal1995} spacecraft) revealed the solar rotation profile down to $0.25\,R_{\odot}$ ($R_{\odot}$ being the radius of the Sun) with an outer radiative core which is rotating as a solid body while strong efforts are still ongoing to constrain the rotation of the nuclear core \citep[e.g.][]{Garciaetal2007,Mathuretal2008}. Space-based asteroseismology with the {\it Kepler} space mission then allowed us to probe the rotation of stars from their surface to their core in the case of main-sequence and evolved low- and intermediate-mass stars \citep[][and references therein]{AertsMathisRogers2019}. First, acoustic modes have been used to study the internal rotation of main-sequence low-mass stars \citep[e.g.][]{Benomaretal2015}. Next, gravito-inertial modes allow us to determine the rotation near the interface of the radiative envelope with the convective core and in the convective core of rapidly-rotating intermediate-mass stars \citep[e.g.][respectively]{VanReethetal2016,Ouazzanietal2020}. Finally, mixed gravito-acoustic modes in evolved low- and intermediate-mass stars provide us precise constrains on the rotation of the internal radiative zone down to the core and the averaged rotation of the envelope \citep[e.g.][]{Mosser2012,Gehanetal2018,Deheuvelsetal2014,Deheuvelsetal2020}. These observations have revealed core-to-surface rotation ratios that are weaker by several orders of magnitude than those predicted by state-of-the-art rotating stellar models for the whole Hertzsprung-Russel diagram \citep[e.g.][]{Eggenbergeretal2012,Ceillieretal2013,Marquesetal2013,Cantielloetal2014,Ouazzanietal2019}. This demonstrates that a powerful mechanism is in action all along the evolution of stars, including our Sun, to extract angular momentum from the deep interior. One of the potential mechanisms is the transport induced by stable \citep[e.g.][]{MestelWeiss1987,MathisZahn2005} or unstable magnetic fields in stably stratified stellar radiation zones \citep[e.g.][]{Spruit2002,Fulleretal2019}. These discoveries have triggered the development of a strong theoretical basis to interpret seismic observations. For oscillation modes, with frequencies far larger than the rotation frequency, the perturbative theory allows us to compute frequency splittings due to the Coriolis and centrifugal accelerations \citep[e.g.][and references therein]{Aertsetal2010,Goupiletal2013}. It also allows us to compute those due to the change of reference frame from the stellar co-rotating one to the observer frame. These splittings are then used to constrain the internal stellar rotation. This method has been intensively used to probe the Solar rotation profile \citep[e.g.][]{Thompsonetal1996,Couvidatetal2003} and the core-to-surface rotation ratio in evolved stars \citep[e.g.][]{Becketal2012,Deheuvelsetal2012,Deheuvelsetal2014,Deheuvelsetal2015,Spadaetal2016,Deheuvelsetal2020}. For low-frequency gravito-inertial modes with frequencies of the same order of magnitude as the rotation frequency, Traditional Approximation of Rotation (TAR), which is not perturbative, can be used \citep[e.g.][]{LeeSaio1997,Townsend2003,VanReethetal2018,MathisPrat2019}\footnote{The Traditional Approximation of Rotation can be used when the buoyancy restoring force dominates the Coriolis force in the direction of the stable entropy and chemical stratification \citep{Mathis2009}.}. Looking at the variation of the period spacing (i.e. the difference between the periods of two consecutive modes) as a function of the modes' period, the (differential) rotation of the oscillations propagation cavity can be constrained \citep[][]{Bouabidetal2013,VanReethetal2018}. This method is currently used to provide key information on the rotation rate of the radiative layers surrounding the convective core in rapidly-rotating intermediate-mass stars \citep[e.g.][]{VanReethetal2016,Christopheetal2018,AertsMathisRogers2019,Lietal2019,Lietal2020}. For each case, the study of the impact of rotation on asymptotic gravity (and gravito-inertial), acoustic, and mixed gravito-acoustic modes, which are rapidly oscillating in the radial direction, has provided both a deep understanding of their modification by the Coriolis (and centrifugal) acceleration and key tools for seismic modelling and inversions.

The second revolution is ground-based high-precision spectropolarimetry that probes the strength and the geometric configuration of magnetic fields at the surface of stars in the whole Hertzsprung-Russell diagram from the pre-main sequence to the late stages of their evolution \citep[e.g.][]{DonatiLandstreet2009,Auriereetal2015,Wadeetal2016}. Large spectropolarimetric surveys thus led our knowledge of dynamo-generated and fossil magnetic fields to a new level of understanding, in particular thanks to the simultaneous development of theoretical models and three-dimensional (3D), global, nonlinear magnetohydrodynamical (MHD) simulations \citep[e.g.][]{BrunBrowning2017,SpruitBraithwaite2017}. However, spectropolarimetry does not give access to the distribution of the magnetic field inside stars. Once again, one has to turn to asteroseismology. The method is the same as the one for rotation, i.e. one searches for specific signatures of the impact of magnetism on the excitation, the propagation, and the damping of stellar oscillations. First studies have considered the effects of simple magnetic configurations like purely dipolar poloidal fields (aligned or inclined respectively to the rotation axis) or purely toroidal fields \citep[e.g.][]{GoodeThompson1992,TakataShibahaschi1994}. However, our knowledge of magnetic topologies and their stability has strongly progressed during the last two decades thanks to the simultaneous development of stellar spectropolarimetry and of 3D MHD simulations. For instance, we now have a better understanding of the formation and the topologies of fossil fields in stellar radiation zones \citep{BraithwaiteSpruit2004}. They should be a combination of dipolar poloidal and toroidal fields to be in their lowest-energy stable equilibrium state \citep{DuezMathis2010}. Therefore, it becomes mandatory to study the seismic signatures of such complex topologies. This has been recently undertaken for the fossil fields in the radiative envelope of upper main-sequence stars \citep{Pratetal2019} and the radiative core of low-mass and intermediate-mass evolved stars (Bugnet et al. 2020) as well as for the dynamo fields generated in the convective envelope of solar-type stars \citep{KieferRoth2018}. 

In this framework, asymptotic analyses have been very useful to provide powerful seismic diagnoses to probe the rotation of stellar interiors \citep{Goupiletal2013}. In this letter, we show how such a method, which has been developed for rotation, can be applied to magnetism. In Sect.~\ref{asymptotic}, we first recall the asymptotic theory for acoustic, gravity, and mixed gravito-acoustic modes. In Sect.~\ref{splittings}, we identify the dominant terms in the magnetic splittings for each class of modes and we derive asymptotic expressions which are the counterpart of those known for stellar rotation. In Sect.~\ref{ProofOfConcept}, we present a proof-of-concept application to a typical intermediate-mass red-giant star. Finally, in Sect.~\ref{conclusion}, we make a synthesis of the results and we discuss how asymptotic predictions can allow us to probe the distribution of the magnetic field at different depths in stars and the applications to different stellar types that can be foreseen.

\section{Asymptotic theory for (mixed) stellar oscillation modes}
\label{asymptotic}
We first recall the basics of the theory of the oscillations of non-rotating and non-magnetic stars, which is necessary to compute first-order magnetic splittings in the asymptotic limits of low-frequency gravity (\emph{g}) modes and high-frequency pressure (\emph{p}) modes. The Lagrangian displacement of an oscillation eigenmode is expanded as 
\begin{equation}
{\vec \xi}\left(\vec r,t\right)={\rm Re}\left\{\left[\xi_{\rm r}\left(r\right)Y_{l}^{m}\left(\theta,\varphi\right){\vec e}_{r}+\xi_{\rm h}\left(r\right){\vec\nabla}_{h}Y_{l}^{m}\left(\theta,\varphi\right)\right]e^{-i\omega t}\right\},
\end{equation}
where $\xi_{\rm r}$ and $\xi_{\rm h}$ are the radial functions of its vertical and horizontal components, $\omega$ is the angular frequency, $Y_l^m$ are the spherical harmonics, $l$ is its degree and $m$ the azimuthal order, ${\vec\nabla}_{h}=\partial_{\theta}\left(\cdot\right){\bf e}_{\theta}+1/\sin{\theta}\,\partial_{\varphi}\left(\cdot\right){\bf e}_{\varphi}$ is the horizontal gradient, $\left\{{\bf e}_r,{\bf e}_{\theta},{\bf e}_{\varphi}\right\}$ are the spherical unit vectors, and ${\rm Re}$ is the real part of a complex number. Following \cite{Gough1993} and \cite{HJCD2017}, who used the adiabatic linearised hydrodynamical equations assuming the Cowling approximation, we derive the Schr\"odinger-like wave equation 
\begin{equation}
\frac{{\rm d}^2 X}{{\rm d}r^2}+k_{\rm r}^{2}\left(r\right)X=0,    
\end{equation}
where $X=c^2\rho^{1/2}{\rm div}\,{\vec\xi}$. The sound speed $c$ is defined as $c=\sqrt{\Gamma_1\,P/\rho}$, where $\rho$ and $P$ are the density and the pressure of the hydrostatic background, respectively, and $\Gamma_{1}=\left(\partial\ln P/\partial\ln \rho\right)_{S}$ is the first adiabatic exponent with $S$ being the macroscopic entropy. The vertical wave number $k_{\rm r}$ is given by
\begin{equation}
k_{\rm r}^{2}=\frac{1}{c^2\left(r\right)}\left[S_{l}^{2}\left(r\right)\left(\frac{N^{2}\left(r\right)}{\omega^2}-1\right)+\omega^2-\omega_{\rm c}^{2}\left(r\right)\right].
\end{equation}
We have identified the Lamb frequency $S_l$ defined by
\begin{equation}
S_{l}^{2}=\frac{l\left(l+1\right)c^2}{r^2}=k_{\rm h}^{2}\,c^2,\quad\hbox{where}\quad k_{\rm h}=\frac{\sqrt{l\left(l+1\right)}}{r}
\end{equation}
is the horizontal wave number, and the Brunt-V\"ais\"al\"a frequency $N$ defined by
\begin{equation}
N^2=g\left(\frac{1}{\Gamma_{1}}\frac{{\rm d}\ln P}{{\rm dr}}-\frac{{\rm d} \ln \rho}{{\rm d}r}\right),
\end{equation}
where $g$ is the gravity of the hydrostatic background. Finally, $\omega_{\rm c}$ is the acoustic cut-off frequency, defined by
\begin{equation}
\omega_{\rm c}^{2}=\frac{c^2}{4H^2}\left(1-2\frac{{\rm d}H}{{\rm d}r}\right),
\end{equation}
where $H=-\left({\rm d}\ln\rho/{\rm d}r\right)^{-1}$ is the density scale height.

\subsection{Asymptotic \emph{g} modes}
In the low-frequency regime for which $\omega\!<\!\!<\!\left\{S_l,N\right\}$, the Lagrangian displacement becomes mostly horizontal (i.e. $\vert\xi_{\rm r}\vert<\!\!<\!\vert\xi_{\rm h}\vert$) and the vertical wave number reduces to \citep[e.g.][]{HJCD2017}
\begin{equation}
k_{\rm r}\approx\frac{N}{\omega}\frac{\sqrt{l\left(l+1\right)}}{r}.
\end{equation}
Using the reduced canonical variable
\begin{equation}
W={\rho}^{1/2}\omega r^2\left(\frac{N^2}{\omega^2}-1\right)^{-1/2}\xi_{\rm h},    
\end{equation}
which has been introduced by \cite{Shibahaschi1979}, and applying the JWKB (for Jeffreys, Wentzel, Kramers, and Brillouin) method \citep[we refer the reader to Appendix \ref{appendix:jwkb} and to][]{Erdalyi1956,FromanFroman2005}, we derive the asymptotic expression for $\xi_{\rm h}\left(r\right)$
\begin{eqnarray}
\lefteqn{\xi_{\rm h}=A_{W}\rho^{-1/2}{\omega}^{-3/2}r^{-3/2}\left[l\left(l+1\right)\right]^{-1/4}N^{1/2}}\nonumber\\
&&\,\times{\rm sin}\,\left(\int_{r_{t;i}}^{r}k_{\rm r}\,{\rm d}r^{'}-\phi^{'}_{g}\right),
\label{AsymptSolG}
\end{eqnarray}
where $A_{W}$ is the amplitude fixed by the excitation and damping mechanisms and the inertia of the mode \citep{Samadietal2015}, $\phi^{'}_{g}$ is a phase, and $r_{t;i}$ ($r_{t;e}$) is the internal (external) turning point for which $k_{\rm r}$ vanishes.

\subsection{Asymptotic \emph{p} modes}
In the high-frequency regime for which $\omega\!>\!>\!S_l$, the Lagrangian displacement becomes mostly vertical (i.e. $\vert\xi_{\rm h}\vert\!<\!<\!\vert\xi_{\rm r}\vert$) and the vertical wave number reduces to
\begin{equation}
k_{\rm r}\approx\frac{\omega}{c}.
\end{equation}
Using the reduced canonical variable
\begin{equation}
V={\rho}^{1/2}c r \left(1-\frac{S_l^2}{\omega^2}\right)^{-1/2}\xi_{\rm r},
\end{equation}
which has been introduced by \cite{Shibahaschi1979}, and applying again the JWKB method (Appendix \ref{appendix:jwkb}), we derive the asymptotic expression for $\xi_{\rm r}\left(r\right)$
\begin{equation}
\xi_{\rm r}=A_{V}{\rho}^{-1/2}c^{-1/2}{\omega}^{-1/2}r^{-1}\cos\left(\int_{r}^{R}k_{\rm r}\,{\rm d}r'-\phi_{p}^{'}\right),
\label{AsymptSolP}
\end{equation}
where $A_{V}$ is the mode's amplitude, $\phi^{'}_{p}$ is a phase, and $R$ is the radius of the star.

\begin{figure*}[t!]
    \centering
    \includegraphics[width=0.45\textwidth]{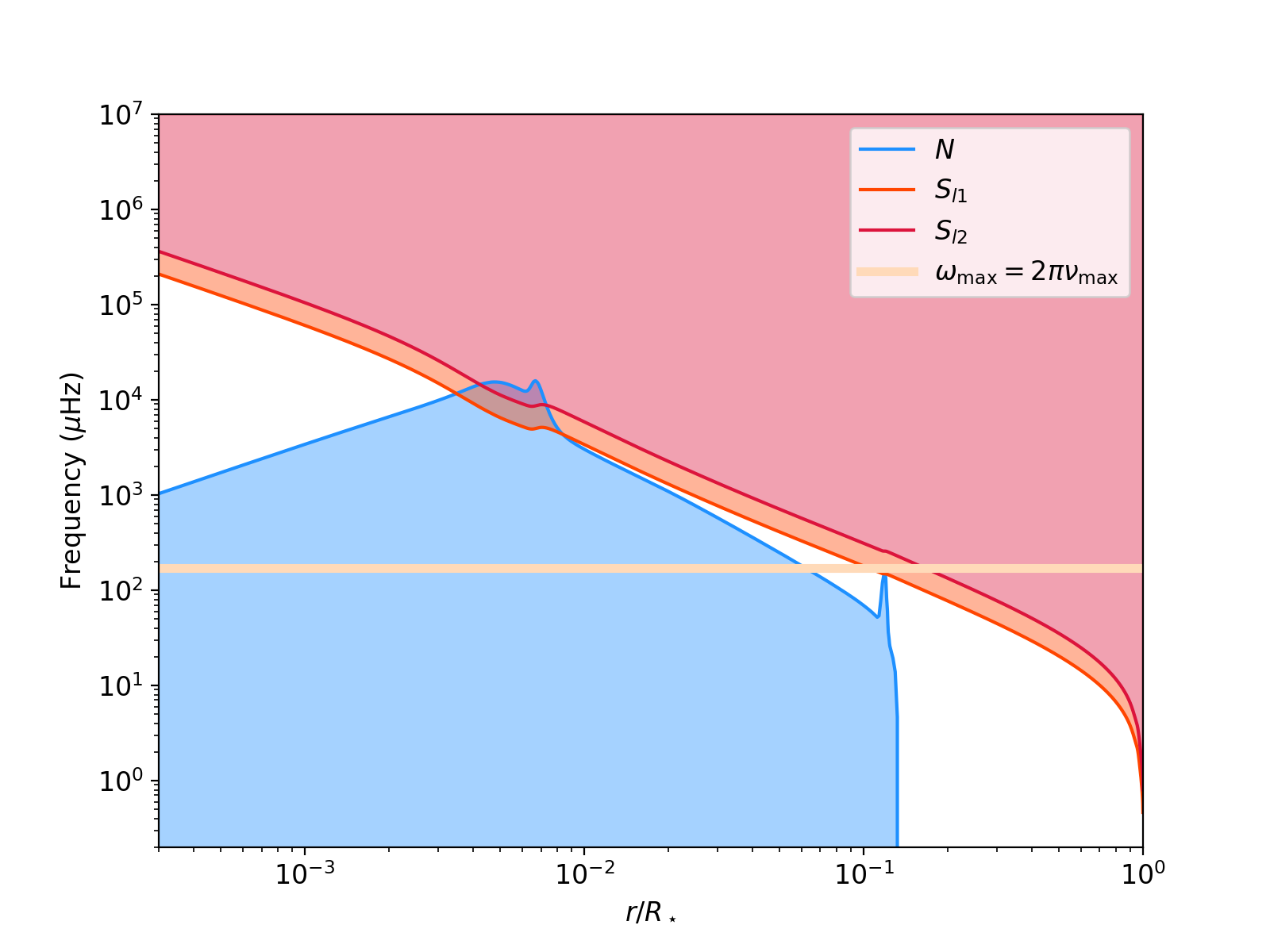}
    \includegraphics[width=0.45\textwidth]{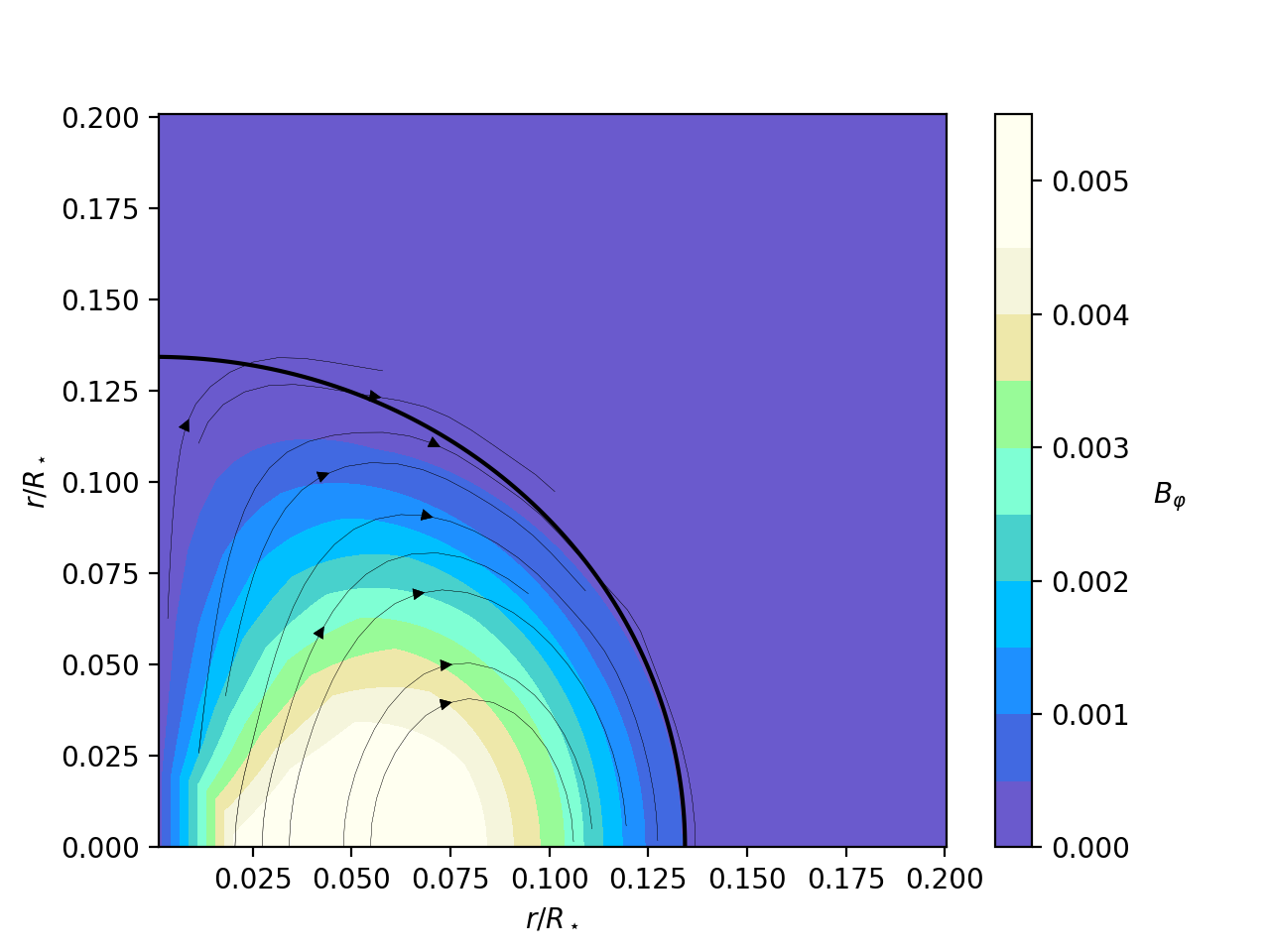}
    \caption{{\bf Left:} Profiles of the Lamb frequencies ($S_l$; red lines) for $l=\left\{1,2\right\}$ and of the Brunt-V\"ais\"al\"a frequency ($N$; blue line) for the studied $1.5M_{\odot}$ $2.8$ Gyr-old intermediate-mass red-giant star (with a solar metallicity $Z=0.02$); the position of $\nu_{\rm max}$ is also reported (orange line). {\bf Right:} Topology of the fossil magnetic field in the stably-stratified radiative core of the studied red giant (its limit is given by the thick black line). The thin black lines are the poloidal field lines, while the intensity of the toroidal field (normalised by the maximum values of the radial field) is given in colors.}
    \label{fig1}
\end{figure*}

\section{Asymptotic magnetic splittings}
\label{splittings}

\subsection{Studied magnetic configuration}
For a sufficiently moderate magnetic field, we can perturb to the first-order the wave equation by the linearised Lorentz force
\begin{equation}
\delta{\vec F}_{\rm L}=\frac{1}{4\pi}\left[\left({\vec\nabla}\wedge{\vec B}\right)\wedge{\delta {\vec B}}+\left({\vec \nabla}\wedge{\delta{\vec B}}\right)\wedge{\vec B}\right],    
\label{LinearisedLorentz}
\end{equation} 
where ${\vec B}$ is the stellar magnetic field, $\delta{\vec B}={\vec\nabla}\wedge\left({\vec \xi}\wedge{\vec B}\right)$
is its fluctuation, and $\mu_0=4\pi$ (in c.g.s.) is the magnetic permeability of vacuum. We obtain the general expression for the magnetic frequency splitting
\begin{equation}
\left(\frac{\delta\omega_{\rm mag}}{\omega_0}\right)=-\frac{\left<{\vec\xi},{\delta{\vec F}_{\rm L}}/\rho\right>}{2\omega_0\left<\vec\xi,\vec\xi\right>},  
\end{equation}
where $
\left<{\vec\xi},{\vec\zeta}\right>=\int_{V}\rho\left({\vec \xi}^{*}\cdot{\vec \zeta}\right){\rm d}V$ (with $^*$ being the complex conjugate) and $\omega_0$ is the angular frequency of the unperturbed mode; we refer the reader to sections 2 \& 3 in \cite{Pratetal2019} for details on the perturbation formalism \footnote{A supplementary term,  ${\vec\nabla}\cdot\left(\rho\boldsymbol{\xi}\right) \left({\vec\nabla}\wedge\mathbf{B}\right)\wedge\mathbf{B}/\left(\mu_0\rho\right)$, should be taken into account in the general case \citep{GoodeThompson1992}. However, we can demonstrate, using again the JWKB approximation, that it is not the dominant term for the asymptotic modes. Indeed, it scales as $k_{\rm r}$ while the dominant terms scales as $k_{\rm r}^2$ while $k_{\rm r}\!>\!\!>\!1$ in the JWKB limit. We assume that the characteristic length scale of the wave is shorter than those of variation of the hydrostatic background and of the magnetic field configuration.}. The mode inertia is expressed as:
\begin{equation}
\langle \boldsymbol{\xi},\boldsymbol{\xi}\rangle = \int_0^R\rho r^2 \left(|\xi_{\rm r}|^2+l\left(l+1\right)|\xi_{\rm h}|^2\right)\textrm{d}r.
\label{eq:massmode}
\end{equation}

In this work, we choose to focus on a dipolar mixed (i.e. with both a poloidal and a toroidal components) axisymmetric configuration
\begin{equation}
    \boldsymbol{B}=B_0\left[ b_{\rm r}(r) \cos{\theta}, b_\theta(r) \sin{\theta}, b_\varphi(r) \sin{\theta}\right],
\end{equation}
where $B_0$ is the field amplitude and the functions $b_{i}\left(r\right)$ (with $i=\left\{r,\theta,\varphi\right\}$) describe the radial dependence of each component. On the one hand, it can model a stable fossil field, which has been formed in a stably stratified radiation zone \citep[e.g.][]{DuezMathis2010}. On the other hand, it can model the first large-scale component of a dynamo-generated magnetic field \citep[e.g.][]{BrunBrowning2017}.

\subsection{Asymptotic \emph{g} modes}
We first focus on low-frequency asymptotic \emph{g} modes, which probe radiative zones, i.e. for low-mass stars their core. For these modes, we have $\omega_0\!<\!\!<\!\left\{S_l,N\right\}$ and thus $k_{\rm r}\approx \left(N/\omega_0\right)k_{\rm h}\!>\!\!>\!k_{\rm h}$ with $k_{\rm h}\equiv\sqrt{l\left(l+1\right)}/r$. Because of their weak compressibility, ${\vec k}\cdot{\vec \xi}\approx 0$ and they are thus quasi-horizontal with $\xi_{\rm v}\!<\!\!<\!\xi_{\rm h}$. Therefore, we select the terms that involve the products $\left\{\xi_{\rm h},\xi_{\rm h}',\xi_{\rm h}''\right\}\times\xi_{\rm h}^{*}$ to isolate the dominant terms in the general expression of the magnetic splitting provided by Eqs. (\ref{FullPoloidal}) \& (\ref{FullToroidal}). Then, using the results provided in Eq. (\ref{AsymptSolG}), we have $\xi_{\rm h}\propto \left(k_{\rm r}\right)^{-1/2}\exp\left[i\int k_{\rm r}{\rm d}r'\right]$ if the studied low-frequency mode is rapidly varying in space in comparison to the characteristic length scale of variation of the stellar hydrostatic structure and of the magnetic configuration. We thus have $\xi_{\rm h}'\propto i k_{\rm r} \xi_{\rm h}$ and $\xi_{\rm h}''\propto k_{\rm r}^2\xi_{\rm h}$. Since $k_{\rm r}H\!>\!\!>\!1$ (since $N\!>\!\!>\!\omega$ and where $H$ is the characteristic length of variation of the hydrostatic background) in the asymptotic regime, the dominant terms are thus those $\propto k_{\rm r}^2$, i.e. $\propto \xi_{\rm h}''\xi_{\rm h}^{*}$. Therefore, as in \cite{Hasanetal2005} and \cite{Rashbaetal2007}, we can identify the dominant terms in the complete expression of the magnetic splitting given in Eq. (\ref{FullPoloidal}). After a last integration by parts, we obtain:
\begin{equation}
\left(\frac{\delta\omega_{\rm mag}}{\omega_0}\right)_g=\frac{B_0^2}{8\pi\omega_0^2}C_{l,m}\frac{\int_{r_{t;i}}^{r_{t;e}}\vert\left(r\,b_{\rm r}\,\xi_{\rm h}\right)'\vert^2\,{\rm d}r}{\int_{r_{t;i}}^{r_{t;e}}\vert\xi_{\rm h}\vert^2{\rho}r^2\,{\rm d}r},    
\label{MagSplitBase}
\end{equation}
where
\begin{equation}
C_{l,m}=\frac{\displaystyle{\int_{0}^{\pi}}\displaystyle{\left[\left|\cos\theta\partial_{\theta}Y_l^m\right|^2+m^2\left|\frac{\cos\theta}{\sin\theta}Y_l^m\right|^2\right]\sin\theta{\rm d}\theta}}{l\left(l+1\right)}.
\label{angularC}
\end{equation}
Since the angular coupling coefficient $C_{l,m}$ is independent of the sign of $m$  in the case of the considered dipolar magnetic configuration, we can already predict that the magnetic frequency splittings for pro- and retrograde modes ($m>0$ and $m<0$, respectively) will be the same. This will allow us to potentially distinguish them from the rotation-induced frequency splittings which depend on the sign of $m$ with opposite values when computed to the first-order in $\Omega$ ($\Omega$ being the rotation of the star); this will be discussed in details in \S \ref{ProofOfConcept}. In the case of main-sequence low-mass stars and of evolved stars, we can assume that $r_{t;i}\approx 0$ and $r_{t;e}\approx R_{\rm core}$ for low-frequency g modes, where $R_{\rm core}$ is the radius of the radiative core. In the case of main-sequence intermediate-mass stars, we have $r_{t;i}\approx R_{\rm CC}$ and $r_{t;e}\approx R$, where $R_{\rm CC}$ is the radius of the convective core.

In addition, we can see that the mostly horizontal low-frequency g modes allows us to probe the orthogonal radial component of the field. This can be understood when looking at the form of the linearised Lorentz force, which implies terms of the form $\vec\xi\wedge{\vec B}$ (cf. Eq. \ref{LinearisedLorentz}). These terms thus couple the mostly horizontal mode's Lagrangian displacement with the orthogonal radial component of the field.

Using Eq. (\ref{AsymptSolG}), it can be written as
\begin{equation}
\left(\frac{\delta\omega_{\rm mag}}{\omega_0}\right)_{g}=\frac{1}{2\omega_0^4}C_{l,m}\frac{\displaystyle{\int_{r_{t;i}}^{r_{t;e}}} \left(\omega_A^r\right)^2 N^2\,\cos^2\left(X_g\right)\,N\,\displaystyle{\frac{{\rm d}r}{r}}}{\displaystyle{\int_{r_{t;i}}^{r_{t;e}}} N \displaystyle{\frac{{\rm d}r}{r}}},
\end{equation}
where $X_g=\int_{r_{t;i}}^{r}k_{\rm r}\,{\rm d}r^{'}-\phi^{'}_{g}$ and $
\omega_A^r=B_0 b_{\rm r}/\left({\sqrt{4\pi\rho}r}\right)$. 

Using the properties of rapidly oscillating integrals (see Appendix \ref{appendix:jwkb}), it reduces to:
\begin{equation}
\left(\frac{\delta\omega_{\rm mag}}{\omega_0}\right)_g=\frac{1}{2}\frac{B_0^2}{4\pi{\rho}_{c}R^2\omega_{0}^{2}}\frac{N_{\rm max}^2}{\omega_0^2} l\left(l+1\right) C_{l,m}\frac{\displaystyle{\int_{x_{t;i}}^{x_{t;e}}}\displaystyle{\frac{b_{\rm r}^2{\widehat N}^2}{\left(\rho/\rho_c\right)x^2}}{\widehat N}\displaystyle{\frac{{\rm d}x}{x}}}{\displaystyle{\int_{x_{t;i}}^{x_{t;e}}}{\widehat N}\displaystyle{\frac{{\rm d}x}{x}}},
\label{AsymptoticIG}
\end{equation}
where we have introduced the dimensionless radius $x=r/R$ (with $x_{t;i}=r_{t;i}/R$ and $x_{t;e}=r_{t;e}/R$), the maximum of the Brunt-V\"ais\"al\"a frequency $N_{\rm max}$ and its dimensionless radial profile ${\widehat N}$ defined such that $N\left(r\right)=N_{\rm max}{\widehat N}(r)$. The form of this integral is really interesting since we recover the one known for asymptotic rotational splittings, but for the radial component of the magnetic field. We can then define a kernel as \cite{Goupiletal2013}  (see the end of Sec. 3.4.) that opens the path for potential inversions when magnetic seismic signatures can be detected (we refer the reader to Bugnet et al. 2020 for the corresponding values of the critical amplitudes).\\

This result can be generalised for g modes with frequencies of the same order of magnitude as the inertial frequency $2\Omega$, where $\Omega$ is the angular velocity of the star. This is the case for instance in rapidly-rotating stars such as $\gamma$ Doradus stars \citep[e.g.][]{VanBeecketal2020} and SPB stars \citep[e.g.][]{Pratetal2019}. In the case where the buoyancy force is larger than the Coriolis force in the direction of the stable entropy or chemical stratification, we can assume the Traditional Approximation of Rotation \citep[TAR; e.g.][]{Eckart1960,LeeSaio1997,Townsend2003,Bouabidetal2013,Mathis2009}. In this approximation, the horizontal projection of the rotation vector is neglected that allows us to separate variables when solving the wave propagation equation as in the non-rotating case. Then, we obtain for g modes modified by rotation, in other words for gravito-inertial modes (hereafter gi modes):   
\begin{eqnarray}
\left(\frac{\delta\omega_{\rm mag}}{\omega_0}\right)_g&=&\frac{1}{2}\frac{B_0^2}{4\pi{\rho}_{c}R^2\omega_{0}^{2}}\frac{N_{\rm max}^2}{\omega_0^2} \Lambda_{k,m}\left(\nu\right){\mathcal C}_{k,m}\left(\nu\right)\nonumber\\
&&\times\frac{\displaystyle{\int_{x_{t;i}}^{x_{t;e}}}\displaystyle{\frac{b_{\rm r}^2{\widehat N}^2}{\left(\rho/\rho_c\right)x^2}}{\widehat N}\displaystyle{\frac{{\rm d}x}{x}}}{\displaystyle{\int_{x_{t;i}}^{x_{t;e}}}{\widehat N}\displaystyle{\frac{{\rm d}x}{x}}},
\label{AsymptoticIGCA}
\end{eqnarray}
with
\begin{equation}
{\mathcal C}_{k,m}\left(\nu\right)=\frac{\displaystyle{\int_{0}^{\pi}\left[H_{\theta}^{2}\left(\cos\theta\right)+H_{\varphi}^{2}\left(\cos\theta\right)\right]\cos^2\theta\sin\theta{\rm d}\theta}}{\displaystyle{\int_{0}^{\pi}\left[H_{\theta}^{2}\left(\cos\theta\right)+H_{\varphi}^{2}\left(\cos\theta\right)\right]\sin\theta{\rm d}\theta}},
\end{equation}
where $\nu=2\Omega/\omega_0$ is the spin parameter and  $\Lambda_{k,m}\left(\nu\right)$ is the horizontal eigenvalues of the Hough functions \citep[e.g.][]{Hough1898,LeeSaio1997,Townsend2003} $\left\{H_r\left(\cos\theta\right), H_{\theta}\left(\cos\theta\right),H_{\varphi}\left(\cos\theta\right)\right\}$ that generalise the spherical harmonics when taking into account the rotation within the TAR (we refer the reader to the Appendix \ref{appen:Hough} for the details of their definition).

\begin{figure*}[t!]
    \centering
    \includegraphics[width=0.45\textwidth]{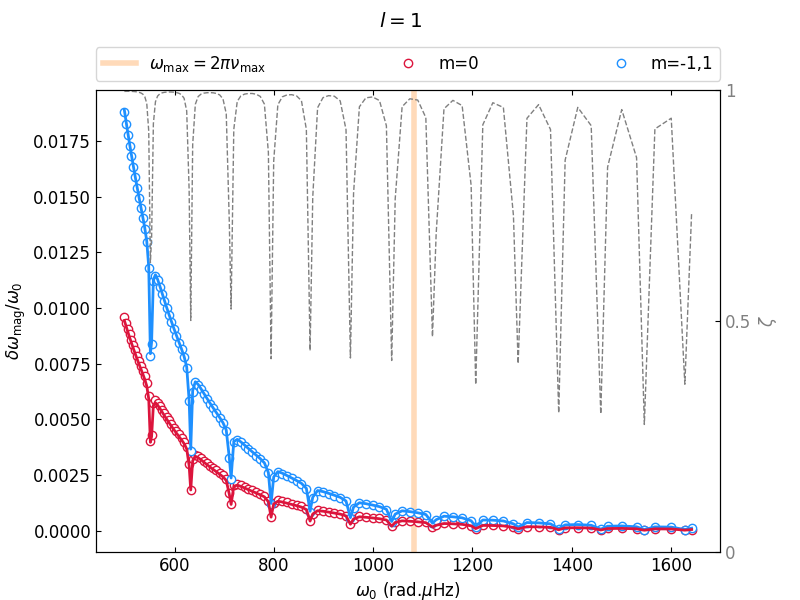}
    \includegraphics[width=0.45\textwidth]{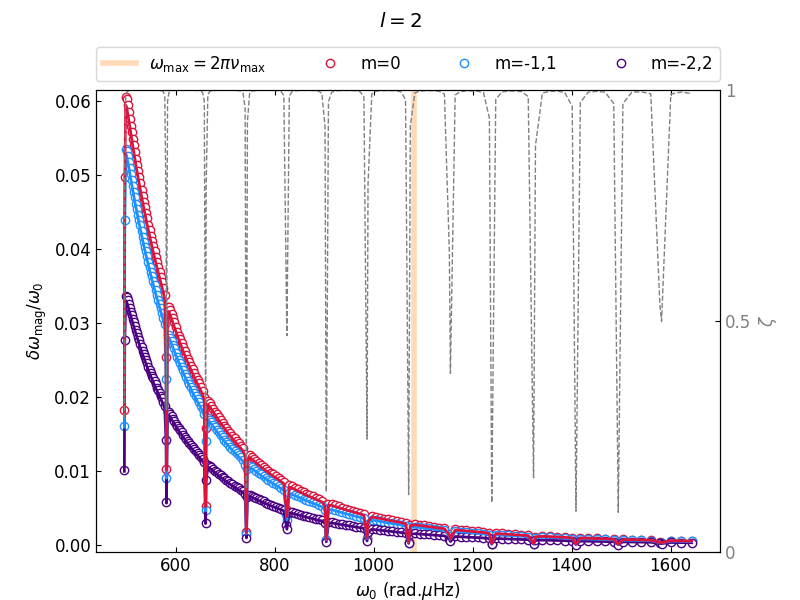}
    \includegraphics[width=0.45\textwidth]{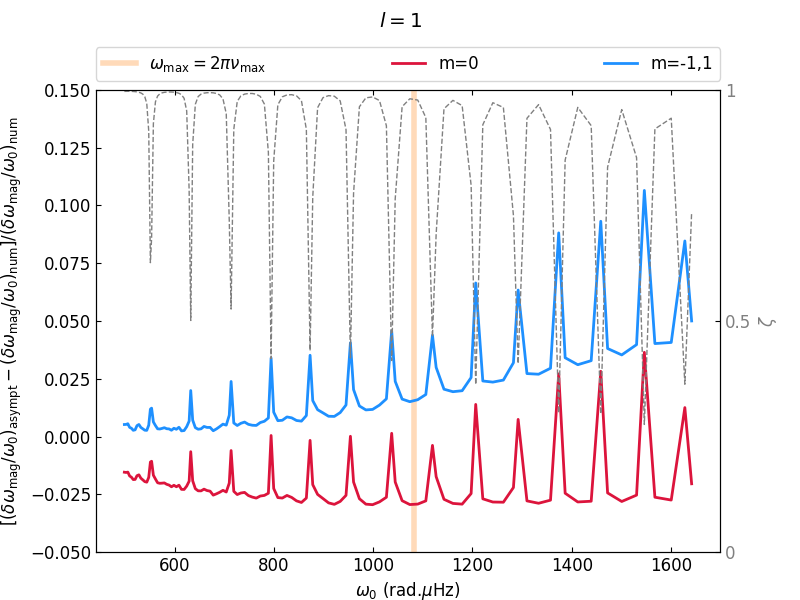}
    \includegraphics[width=0.45\textwidth]{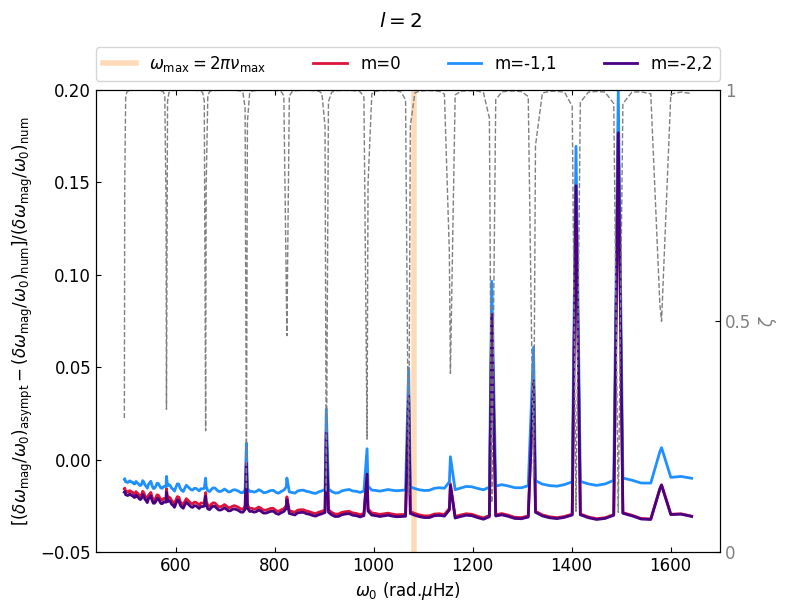}
    \caption{First-order magnetic splitting for dipolar ($l=1$, left-top panel) and quadrupolar ($l=2$, right-top panel) mixed modes in the studied $1.5\,M_{\odot}$ red giant star. The dots correspond to the full expression for the splitting (eqs. \ref{FullPoloidal} and \ref{FullToroidal}) computed numerically using the GYRE and MESA codes while the solid lines are the asymptotic solutions given in Eq. (\ref{AsymptoticIG}). The $\zeta$ function related to the mixed character of modes propagating in evolved low-mass stars is plotted in grey dotted line. Finally, the vertical thick orange line indicates the position of $\omega_{\rm max}=2\pi\nu_{\rm max}$. Bottom panels: relative errors between the asymptotic magnetic splittings as derived in Eq. \ref{AsymptoticIG} and their full expression given in Appendix \ref{appen:generalsplitting} for dipolar (left panel) and quadrupolar (right panel) modes, respectively.}
    \label{fig2}
\end{figure*}

\subsection{Asymptotic \emph{p} modes}
We can apply the same method to the case of high-frequency acoustic modes, which probe the external layers of stars. In the same way, we isolate using Eq. (\ref{AsymptSolP}) the dominant terms in Eq. (\ref{FullToroidal}), which are $\propto \xi_{\rm r}''\xi_{\rm r}^{*}$
\begin{equation}
\left(\frac{\delta\omega_{\rm mag}}{\omega_0}\right)_p=\frac{B_0^2}{8\pi\omega_0^2}D_{l,m}\sum_{i=\left\{\theta,\varphi\right\}}\frac{\int_{r_{t}}^{R}\vert\left(r\,b_i\,\xi_{\rm r}\right)'\vert^{2}\,{\rm d}r}{\int_{r_{t}}^{R}\vert\xi_{\rm r}\vert^2{\rho}r^2\,{\rm d}r},
\end{equation}
where $D_{l,m}=\int_{0}^{\pi}\sin^2\theta\vert Y_l^m\vert^2\sin\theta{\rm d}\theta$ is computed in Appendix (\ref{Dcompute}). As for the angular coupling coefficients $C_{l,m}$ computed for asymptotic g modes, we also identify here that $D_{l,-m}=D_{l,m}$. We have introduced the internal turning point $r_{t}=\sqrt{l\left(l+1\right)}c/\omega_0$ for which $S_l\left(r_{t}\right)=\omega_{0}$. In the case of low-degree high-frequency acoustic modes, we have $r_{t}\rightarrow 0$. Using the asymptotic JWKB solution given in Eq. (\ref{AsymptSolP}) and the theory of rapidly oscillating integrals (see Appendix \ref{appendix:jwkb}), we finally get:      
\begin{equation}
\left(\frac{\delta\omega_{\rm mag}}{\omega_0}\right)_{p}=\frac{D_{l,m}}{2}\sum_{i=\left\{\theta,\varphi\right\}}\frac{\displaystyle{\int_0^{R}\frac{\left(V_{\rm A}^{i}\right)^{2}}{c^2}\frac{{\rm d}r'}{c}}}{\displaystyle{\int_0^{R}\frac{{\rm d}r'}{c}}},
\label{AcousticFinal}
\end{equation}
where $V_{\rm A}^{i}=B_i/{\sqrt{4\pi\rho}}$. We recover again exactly the same form as the one for the asymptotic rotational splittings, but for the horizontal components of the magnetic field.

As in the case of low-frequency g modes, we can note that high-frequency acoustic modes, which have mostly vertical Lagrangian displacements, allow us to probe the orthogonal horizontal (latitudinal and azimuthal) components of the field.     

\subsection{Mixed modes}
As in \cite{Goupiletal2013}, we can express the magnetic splittings for mixed (gravito-acoustic) modes using the expressions of those of asymptotic gravity and acoustic modes:
\begin{equation}
\left(\frac{\delta\omega_{\rm mag}}{\omega_0}\right)=\left(\frac{\delta\omega_{\rm mag}}{\omega_0}\right)_{g}\zeta+\left(\frac{\delta\omega_{\rm mag}}{\omega_0}\right)_{p}\left(1-\zeta\right),
\end{equation}
where we have introduced the 
\begin{equation}
\zeta=\frac{I_g}{I}=
 \frac{\int_{r_{t;i}}^{r_{t;e}} \left(\xi_{\rm r}^2+l\left(l+1\right)\xi_{\rm h}^2\right) r^2 dr}{\int_0^{R} \left(\xi_{\rm r}^2+l\left(l+1\right)\xi_{\rm h}^2\right) r^2 dr}    
\end{equation}
function with $I_g$ and $I$ being the inertia of their g-dominated component (the so-called g-m modes) and their total inertia, respectively. It quantifies their behavior as g-dominated mode when $\zeta=1$ or as p-dominated mode (the so-called p-m modes) when $\zeta\rightarrow 0$.

It is interesting here to make the analogy with the case of rotation studied in \cite{Goupiletal2013}. They showed that for the dipolar modes ($l=1$)
\begin{equation}
\delta\omega\approx\frac{\zeta}{2}\left<{\Omega}\right>_{\rm core}+\left(1-\zeta\right)\left<{\Omega}\right>_{\rm envelope},    
\end{equation}
where
\begin{equation}
\left<{\Omega}\right>_{\rm core}\approx\frac{1}{2\sqrt{l\left(l+1\right)}}\left(\frac{\omega_0}{\Omega_c}\right)^2\frac{\displaystyle{\int_{\rm core}\Omega\left(x\right)N\frac{{\rm d}x}{x}}}{\displaystyle{\int_{\rm core}N\frac{{\rm d}x}{x}}},
\end{equation}
with $\Omega_c=(GM/R^3)^{1/2}$ the critical Keplerian velocity, and
\begin{equation}
\left<{\Omega}\right>_{\rm envelope}\approx\frac{1}{2}\frac{\displaystyle{\int_{\rm envelope}\Omega\left(x\right)\frac{{\rm d}x}{c}}}{\displaystyle{\int_{\rm envelope}\frac{{\rm d}x}{c}}}.
\end{equation}
Using Eqs. (\ref{AsymptoticIG}) and (\ref{AcousticFinal}), we identify the analogy between $\Omega$ in the core and $\left(\omega_A^r\right)^2 N^2/\omega_0^3$ in the case of g-dominated modes and between $\Omega$ in the envelope and $\omega_0 \left(V_{\rm A}^{i}\right)^2/c^2$ for p-dominated modes. 

This shows the potential power of mixed modes to probe the magnetism of stars where they propagate from their surfaces to their cores.

\begin{figure*}[t!]
    \centering
    \includegraphics[width=0.95\textwidth]{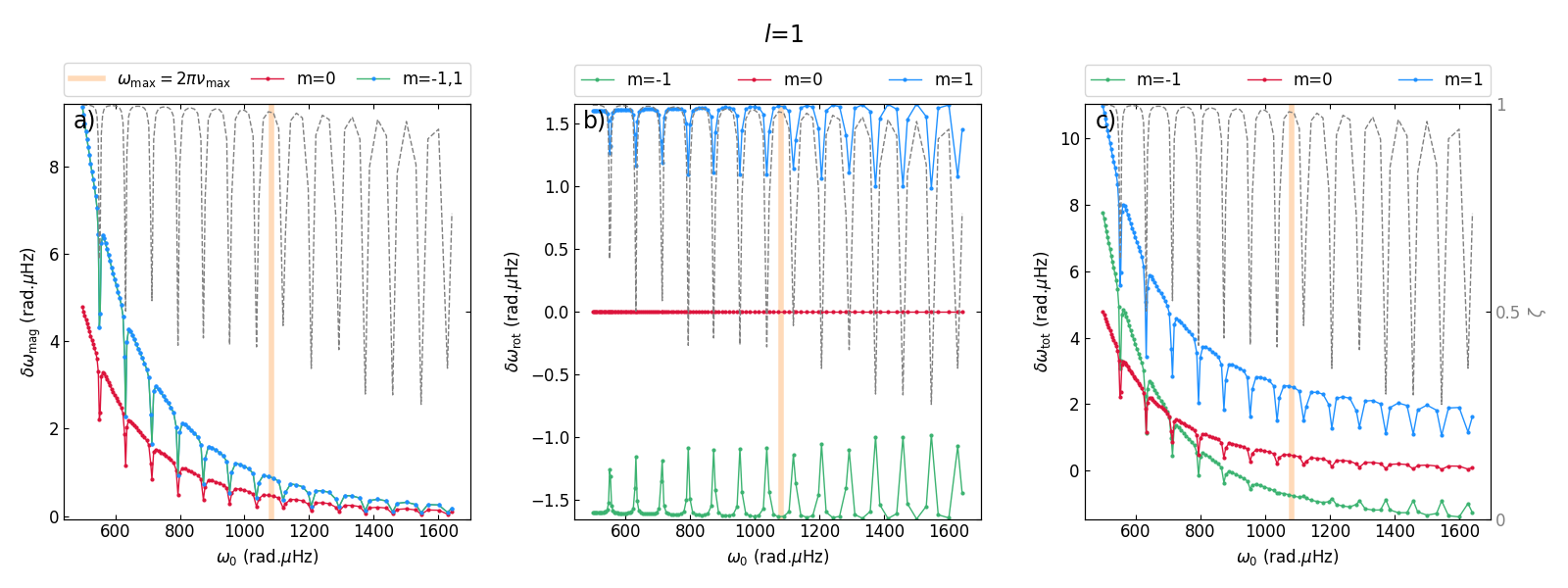}
    \includegraphics[width=0.95\textwidth]{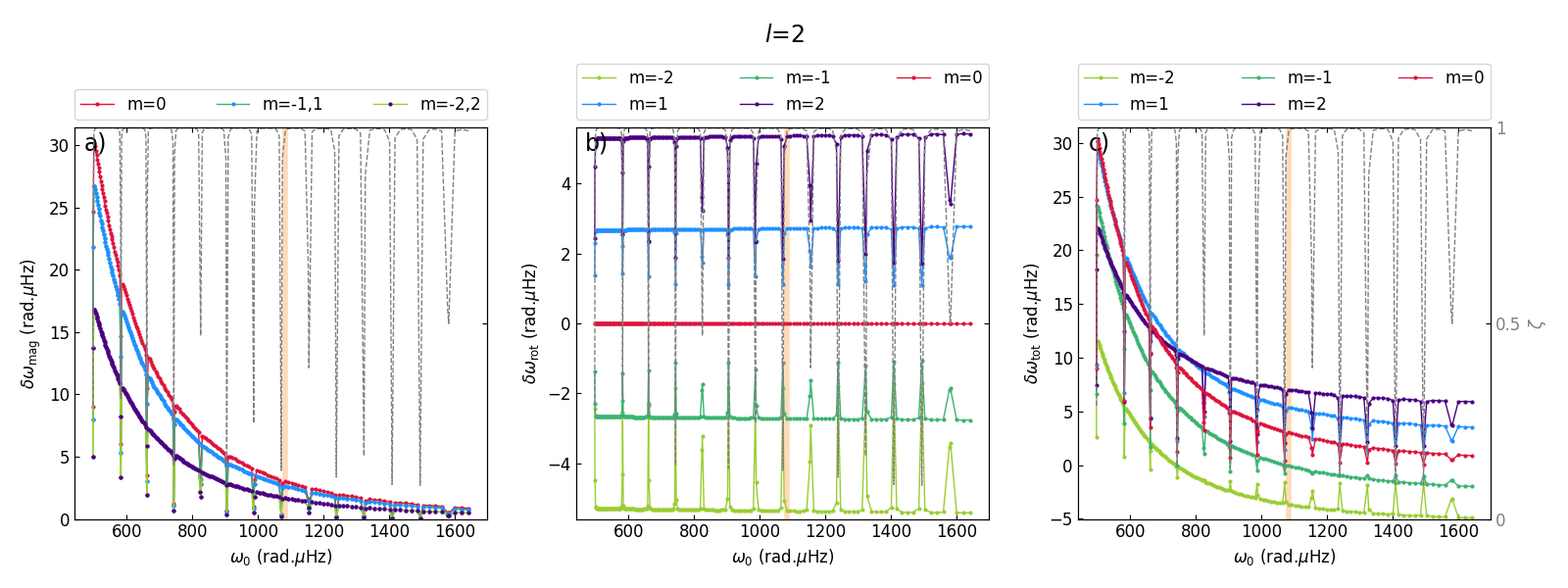}
    \caption{Leading order perturbations on the frequencies of dipolar ($l=1$, top panels) and quadripolar ($l=2$, bottom panels) mixed modes in the studied $1.5\,M_{\odot}$ red giant star. Left panels represent the magnetic angular frequency splitting $\delta\omega_{\rm mag}$ (for $B_0=1$MG) of the $m \in [-l, l]$ modes. On middle panels the angular frequency splitting $\delta\omega_{\rm rot}$ for a plausible two-layer differential rotation (with $\Omega_{\mathrm{core}}= 0.5\,\mu{\rm Hz}$ and $\Omega_{\mathrm{env}}=  \Omega_{\mathrm{core}}$/10) of the same mixed modes is represented. Finally, right panels show the combination of magnetic and rotational splittings of mixed-mode angular frequencies $\delta\omega_{\rm tot}=\delta\omega_{\rm mag}+\delta\omega_{\rm rot}$. The orange vertical line indicates on each panel the expected frequency of maximum power of the modes $\nu_{\mathrm{max}}$. The $\zeta$ function is represented in each case by the grey dashed line.}
    \label{fig3}
\end{figure*}

\section{A key application: the case of evolved low-mass and intermediate-mass stars}
\label{ProofOfConcept}

Because of all the key results they have provided to the theory of stellar evolution and to the study of the internal dynamics of stars thanks to space-based asteroseismology \citep[e.g.][]{HJCD2017,AertsMathisRogers2019}, we choose here to apply our theoretical results to a typical $1.5M_{\odot}$ $2.8$ Gyr-old intermediate-mass red-giant star (with a solar metallicity $Z=0.02$). The profiles of its Brunt-Va\"is\"al\"a and Lamb frequencies for dipolar and quadrupolar modes ($N$, $S_{l=1}$, $S_{l=2}$, respectively) are represented in Fig. \ref{fig1} (left panel) where the position of $\nu_{\rm max}=172.2$ $\mu$Hz, the frequency of maximum oscillation power, is reported by the thick horizontal orange line.

First, we compute the topology of a mixed (with both poloidal and toroidal components) relaxed fossil magnetic field in its radiative core following the method introduced in \cite{Pratetal2019} and the theoretical modelling for such field by \cite{DuezMathis2010}. As in Bugnet et al. (2020), the chosen boundary conditions are $b_{\rm r}=0$ at the radiative core/convective envelope boundary. Its amplitude is fixed to $B_0=10^6\,$G and its geometry is reported in Fig. \ref{fig1} (right panel). For such a possible field amplitude \citep{Cantielloetal2016}, magnetic frequency splittings should be detectable within {\it Kepler} data (Bugnet et al., submitted). The black thin lines represent the poloidal field lines while the colors give the amplitude of the toroidal component of the field (the boundary of the radiative core is given by the thick black line).

Then, we compute the magnetic splittings with their complete expression $\left(\delta\omega_{\rm mag}/\omega_0\right)_{\rm num}$ given in Appendix \ref{appen:generalsplitting} using the combination of the GYRE and MESA codes \citep[][]{Townsendetal2013,Paxtonetal2011} for the dipolar ($l=1$) and quadrupolar ($l=2$) mixed modes (represented by the dots in the left panel (resp. right panel) of Fig. \ref{fig2}). We compute the corresponding asymptotic prediction $\left(\delta\omega_{\rm mag}/\omega_0\right)_{\rm asympt}$ using Eq. (\ref{AsymptoticIG}) (in solid lines in Fig. \ref{fig2}). The complete and asymptotic expressions are compared thanks to the relative error defined as $\left[\left(\delta\omega_{\rm mag}/\omega_0\right)_{\rm asympt}-\left(\delta\omega_{\rm mag}/\omega_0\right)_{\rm num}\right]/\left(\delta\omega_{\rm mag}/\omega_0\right)_{\rm num}$ (Fig. \ref{fig2}, bottom panels). In the low-frequency regime, the agreement is excellent for g-dominated mixed modes with a relative error lower than {\bf $4\%$} for dipolar modes and {\bf $3\%$} for quadrupolar modes outside the dips of the $\zeta$ function. These results are coherent with the high radial orders of the obtained mixed modes, i.e. $n\in\left[-138,-24\right]$ for the dipolar modes and $n\in\left[-244,-56\right]$ for the quadrupolar ones, for which the JWKB approximation works well. This demonstrates the high potential of using the asymptotic magnetic splittings for intensive seismic modelling where they can be very useful to explore a broad space of stellar parameters \citep[see e.g. the work by][for main-sequence intermediate-mass stars]{VanBeecketal2020}.    

As this has already been identified when looking at Eq. (\ref{angularC}), the effects of the studied dipolar mixed (poloidal + toroidal) fossil magnetic field on the frequencies of g-dominated mixed modes is the same for pro- and retrograde modes. This strongly differs from the case of the frequency splittings induced by the stellar rotation, which are due to the combination of the Coriolis acceleration and of the change of reference frame \citep[e.g.][]{Aertsetal2010}: 
\begin{equation}
\delta\omega_{\rm rot}=-\frac{\langle \boldsymbol{\xi},\boldsymbol{F_c}(\boldsymbol{\xi})\rangle + \langle \boldsymbol{\xi},\boldsymbol{F_f}(\boldsymbol{\xi}) \rangle}{2 \omega_0 \langle \boldsymbol{\xi}, \boldsymbol{\xi} \rangle},
    \label{eq:w1}
\end{equation}
where $\boldsymbol{F_{c}}(\boldsymbol{\xi})=2\iu\omega_0\boldsymbol{\Omega} \wedge \boldsymbol{\xi}$ is the Coriolis acceleration operator and $\boldsymbol{F_{f}}(\boldsymbol{\xi})= -2 m\omega_0\Omega\boldsymbol{\xi}$ is the advection operator due to the change of reference frame. In the case of a radial differential rotation, we have:
\begin{equation}
\langle \boldsymbol{\xi},\boldsymbol{F_c}(\boldsymbol{\xi}) \rangle = 4 m \omega_0  \int_0^R \rho r^2 \Omega(r)\left[|\xi_{\rm h}|^2+2\xi_{\rm r}^* \xi_{\rm h}\right]\textrm{d}r \\ 
\label{eq:corioliseffect}
\end{equation}
and
\begin{equation}
\langle \boldsymbol{\xi},\boldsymbol{F_f}(\boldsymbol{\xi}) \rangle~=~{-2} m \omega_0
\left(\int_0^{R} \rho r^2 \left(|\xi_{\rm r}|^2+l\left(l+1\right)|\xi_{\rm h}|^2\right) \Omega(r) \textrm{d}r \right).
\label{eq:frameeffect}
\end{equation}
Therefore, $\delta\omega_{\rm rot}$ is proportional to $m$ leading to opposite rotational frequency splittings for pro- and retrograde modes. This is illustrated in Fig. \ref{fig3}, where we represent in the middle column this rotational splitting when assuming a typical two-zone differential rotation with a solid-body rotation $\Omega_{\rm core}/\left(2\pi\right)=0.5\,\mu {\rm Hz}$ in the radiative core and $\Omega_{\rm envelope}/\left(2\pi\right)=\left(\Omega_{\rm core}/10\right)/2\pi=0.05\,\mu{\rm Hz}$ in the convective envelope \citep{Gehanetal2018}. When added to the magnetic frequency splitting, which is the same for pro- and retrograde modes, this leads to an asymmetry of the total frequency splitting that scales with the squared amplitude of the field (cf. Eqs. \ref{MagSplitBase} \& \ref{AsymptoticIG}). This asymmetry will allow asteroseismologists to probe the potential presence of an (axisymmetric) stable fossil magnetic field in the core of evolved low-mass and intermediate-mass stars. We refer the reader to Bugnet et al. (submitted) for a detailed characterisation in seismic data of this asymmetry.\\

We do not consider here the case of p-dominated modes during the sub-giant phase since Bugnet et al. (2020) demonstrated that magnetic signatures would not be detectable for them.

\section{Conclusion and perspectives}
\label{conclusion}

In this work, we have established the asymptotic values of the magnetic splittings of low-frequency gravity and gravito-inertial (\emph{gi}) modes and high-frequency acoustic modes. These results have been applied to the case of mixed gravito-acoustic modes as those that propagate in evolved low- and intermediate-mass red-giant stars. They have been derived in the case of a large-scale axisymmetric mixed (poloidal + toroidal) dipolar field that can be representative of a stable fossil field or of the large-scale axisymmetric dipolar component of a dynamo-generated field. Main results that we have obtained are the following:
\begin{itemize}
    \item the expressions obtained for the magnetic splittings in the asymptotic limits of low-frequency \emph{g} (\emph{gi}) modes and high-frequency \emph{p} modes are very similar in their form to those of rotational splittings in the same limits. This opens the path to potential inversions of the internal distribution of magnetic fields in stellar interiors when those signatures can be detected in stellar oscillation frequency spectra; 
    \item For asymptotic \emph{g} (\emph{gi}) modes, the splittings scale as the normalised squared ratio of the local Alfv\'en frequency to the mode frequency integrated along the buoyancy radius. For asymptotic \emph{p} modes, they scale as the normalised squared ratio of the Alfv\'en velocity to the sound speed integrated along the acoustic path; 
    \item in the case of the studied dipolar configuration, magnetic frequency splittings are the same for pro- and retrograde modes ($m>0$ and $m<0$, respectively). This allows us to distinguish them from rotational frequency splittings that have opposite values for pro- and retrograde modes. This leads to an asymmetry of the total frequency splittings when adding the two effects. It could be detectable within high-precision asteroseismic data for sufficient field amplitude. We refer the reader to Bugnet et al. (submitted) for a complete seismic characterisation of this asymmetry and for the study of such a needed amplitude along the evolution of low-mass and intermediate-mass stars;
    \item In the case of the large-scale mixed dipolar configuration, which is treated here, low-frequency \emph{g} (\emph{gi}) modes and high-frequency \emph{p} modes are sounding different components of the magnetic field because of the form of the linearised Lorentz force which involves $\vec\xi\wedge\vec B$ terms. On the one hand, mostly horizontal low-frequency \emph{g} (\emph{gi}) modes, which propagate in the central regions of stars, are probing the orthogonal radial component of the field. On the other hand, mostly-vertical high-frequency \emph{p} modes, which propagate in the external regions of stars, are probing the orthogonal horizontal (latitudinal and azimuthal) components of the field. As in the case of the rotation, where mixed modes have allowed us to probe the rotation of the core and of the external envelope thanks to their \emph{g}-dominated (g-m) and \emph{p}-dominated (p-m) components, respectively, this opens the path to probe the stellar magnetic field from the surface to the core of stars. Since high frequencies of asymptotic \emph{p} modes will be more distant than low-frequencies of asymptotic \emph{g} modes from the Alfv\'en frequency, the seismic signatures of the external field can be more difficult to detect in the case of low-amplitude field. This can be fixed in the case of bright stars as those observed by TESS (Transiting Exoplanet Survey Satellite) \citep{Rickeretal2015}, and in the future by the PLATO (PLAnetary Transits and Oscillations of stars) mission \citep{Raueretal2014}, by combining spectropolarimetic observations to determine the configuration and the strength of the field at the surface and seismic signatures to probe the core magnetism. This strategy is summarised in Fig. \ref{Strategy}.
    \begin{figure}[h!]
    \centering
    \includegraphics[width=0.45\textwidth]{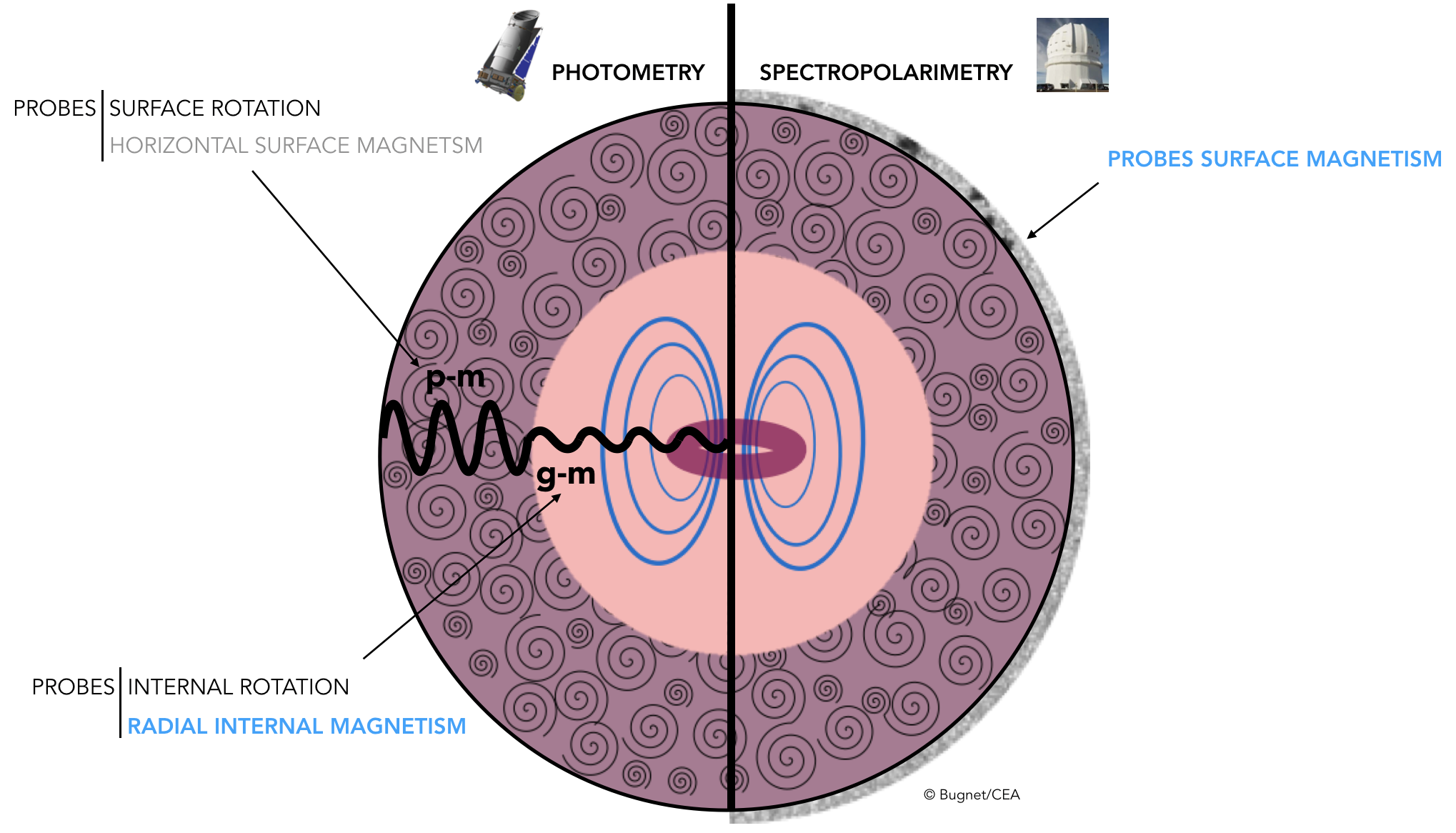}
    \caption{Global strategy to probe the internal magnetism of stars using asteroseismology (in potential synergy with spectropolarimetry). Here, the scheme is designed for evolved low- and intermediate-mass stars.}
    \label{Strategy}
    \end{figure}
\end{itemize}

In a near future, the derived asymptotic relations should be tested and used systematically for all classes of pulsators. In addition, more complex (non-dipolar/non-axisymmetric) magnetic topologies must be studied \citep[e.g.][]{Pratetal2020}. This should allow us to build step by step a complete knowledge of the magnetism of stars from their surfaces to their cores. 
\begin{acknowledgements}
The authors thank the referee and Pr. J. Christensen-Dalsgaard for their very constructive comments and remarks that allowed us to improve the article. St. M., L. B., V. P., and K. A. acknowledge support from the European Research Council through ERC grant SPIRE 647383. All the members from CEA acknowledge support from GOLF and PLATO CNES grants of the Astrophysics Division at CEA. S. Mathur acknowledges support by the Ramon y Cajal fellowship number RYC-2015-17697. We made great use of the megyr python package for interfacing MESA and GYRE codes. 
\end{acknowledgements}    
\bibliographystyle{aa} 
\bibliography{bibMB}

\begin{appendix}


\section{Hough functions}
\label{appen:Hough}

The radial Hough functions \citep{Hough1898} are defined by $H_r(\cos\theta) = f(\mu=\cos\theta)$, where $f$ is the solution of the so-called Tidal Laplace Equation
\begin{eqnarray}
    \lefteqn{\frac{1-\mu^2}{1-\nu^2\mu^2}\frac{{\rm d}^2f}{{\rm d}\mu^2} - \frac{2\mu(1-\nu^2)}{(1-\nu^2\mu^2)^2}\frac{{\rm d}f}{{\rm d}\mu}}\nonumber\\
    &+&\left[\frac{m\nu(1+\nu^2\mu^2)}{(1-\nu^2\mu^2)^2}-\frac{m^2}{(1-\mu^2)(1-\nu^2\mu^2)}\right]f\nonumber\\
    &=&-\Lambda\left(\nu\right) f,
\end{eqnarray}
where $\nu=2\Omega/\omega$ is the spin factor.
In the non-rotating case ($\nu=0$), the eigenvalue $\Lambda$ \citep[they are generally indexed by an integer $k$ for a given $m$; see e.g.][]{LeeSaio1997} reduces to $l(l+1)$, where $l$ is the angular degree of the mode, and $H_r$ simplifies to the classical associated Legendre polynomial $P_{l}^m\left(\cos\theta\right)$.

The latitudinal and azimuthal Hough functions are respectively derived:
\begin{equation}
    H_\theta\left(\cos\theta\right) \sin\theta = \frac{H_r' \sin\theta - m\nu H_r\cos\theta}{1-\nu^2\cos^2\theta}
\end{equation}
and
\begin{equation}
    H_\varphi\left(\cos\theta\right) \sin\theta = \frac{mH_r-\nu H_r'\sin\theta\cos\theta}{1-\nu^2\cos^2\theta},
\end{equation}
where $'$ is the total latitudinal derivative. $H_\varphi$ has the same parity as $H_r$ with respect to $\theta=\pi/2$, while $H_\theta$ has the opposite parity. In the non-rotating case ($\nu=0$), $H_{\theta}$ and $H_{\varphi}$ reduce to ${\rm d}P_l^m/{\rm d}\theta$ and $m P_l^m/\sin\theta$, respectively.

\section{General expression of magnetic splittings}
\label{appen:generalsplitting}

Here, we recall all the non-zero-average terms of the work of the Lorentz force $\mu_0\delta\vec F_{\rm L}\cdot\vec\xi^*$.
All these terms involve either only poloidal components of the magnetic field, or only the toroidal component, and they are grouped accordingly as in \cite{Pratetal2019}.
All terms are composed by a purely radial part multiplied by a purely latitudinal part, and the prime symbol ($'$) is a total radial or latitudinal derivative, depending on the considered part.

\subsection{Poloidal terms}

We define $A = [(rb_\theta)'+b_{\rm r}]$. The terms that involve poloidal components are obtained:
\begin{eqnarray}
    &&-m\frac{\xi_{\rm h}b_\theta A\xi_{\rm r}^*}{r^2}H_{\rm r}H_\varphi\sin\theta
    +\frac{(r\xi_{\rm r}b_\theta)'A\xi_{\rm r}^*}{r^2}H_{\rm r}^2\sin^2\theta\nonumber\\
&&    -\frac{(r\xi_{\rm h}b_{\rm r})'A\xi_{\rm r}^*}{r^2}H_{\rm r}H_\theta\sin\theta\cos\theta
    +\frac{\xi_{\rm r}b_\theta A\xi_{\rm h}^*}{r^2}H_\theta(H_{\rm r}\sin^2\theta)'\nonumber\\
&& -\frac{Ab_{\rm r}|\xi_{\rm h}|^2}{r^2}H_\theta(H_\theta\sin\theta\cos\theta)'
    +m\frac{Ab_{\rm r}|\xi_{\rm h}|^2}{r^2}H_\theta H_\varphi\cos\theta\nonumber\\
&&-m\frac{(\xi_{\rm h}b_\theta)'b_\theta\xi_{\rm r}^*}{r}H_{\rm r}H_\varphi\sin\theta
    +\frac{(r\xi_{\rm r}b_\theta)''b_\theta\xi_{\rm r}^*}{r}H_{\rm r}^2\sin^2\theta\nonumber\\
&&-\frac{(r\xi_{\rm h}b_{\rm r})''b_\theta\xi_{\rm r}^*}{r}H_{\rm r}H_\theta\sin\theta\cos\theta
    +\frac{b_\theta^2|\xi_{\rm r}|^2}{r^2}H_{\rm r}\sin\theta\left[\frac{(H_{\rm r}\sin^2\theta)'}{\sin\theta}\right]'\nonumber\\
&&-\frac{\xi_{\rm h}b_{\rm r}b_\theta\xi_{\rm r}^*}{r^2}H_{\rm r}\sin\theta\left[\frac{(H_\theta\sin\theta\cos\theta)'}{\sin\theta}\right]'\nonumber\\
&&+m\frac{\xi_{\rm h}b_{\rm r}b_\theta\xi_{\rm r}^*}{r^2}H_{\rm r}\sin\theta\left(H_\varphi\frac{\cos\theta}{\sin\theta}\right)'
    +m\frac{(\xi_{\rm h}b_\theta)'b_{\rm r}\xi_{\rm h}^*}{r}H_\theta H_\varphi\cos\theta\nonumber\\
&&-\frac{(r\xi_{\rm r}b_\theta)''b_{\rm r}\xi_{\rm h}^*}{r}H_{\rm r}H_\theta\sin\theta\cos\theta
    +\frac{(r\xi_{\rm h}b_{\rm r})''b_{\rm r}\xi_{\rm h}^*}{r}H_\theta^2\cos^2\theta\nonumber\\
&&-\frac{\xi_{\rm r}b_{\rm r}b_\theta\xi_{\rm h}^*}{r^2}H_\theta\cos\theta\left[\frac{(H_{\rm r}\sin^2\theta)'}{\sin\theta}\right]'\nonumber\\
&&+\frac{b_{\rm r}^2|\xi_{\rm h}|^2}{r^2}H_\theta\cos\theta\left[\frac{(H_\theta\sin\theta\cos\theta)'}{\sin\theta}\right]'\nonumber\\
&&-m\frac{b_{\rm r}^2|\xi_{\rm h}|^2}{r^2}H_\theta\cos\theta\left(H_\varphi\frac{\cos\theta}{\sin\theta}\right)'
    +\frac{(r\xi_{\rm h}b_{\rm r})'b_\theta\xi_{\rm h}^*}{r^2}H_\varphi(H_\varphi\sin\theta\cos\theta)'\nonumber\\
&&+\frac{b_\theta^2|\xi_{\rm h}|^2}{r^2}H_\varphi[\sin\theta(H_\varphi\sin\theta)']'
    -m^2\frac{b_\theta^2|\xi_{\rm h}|^2}{r^2}H_\varphi^2\nonumber\\
&&+m\frac{(r\xi_{\rm r}b_\theta)'b_\theta\xi_{\rm h}^*}{r^2}H_{\rm r}H_\varphi\sin\theta
    -m\frac{(r\xi_{\rm h}b_{\rm r})'b_\theta\xi_{\rm h}^*}{r^2}H_\theta H_\varphi\cos\theta\nonumber\\
&&-m\frac{\xi_{\rm r}b_\theta b_{\rm r}\xi_{\rm h}^*}{r^2}H_\varphi\frac{\cos\theta}{\sin^2\theta}(H_{\rm r}\sin^2\theta)'
    +m\frac{b_{\rm r}^2|\xi_{\rm h}|^2}{r^2}H_\varphi\frac{\cos\theta}{\sin^2\theta}(H_\theta\sin\theta\cos\theta)'\nonumber\\
&&-m^2\frac{b_{\rm r}^2|\xi_{\rm h}|^2}{r^2}H_\varphi^2\frac{\cos^2\theta}{\sin^2\theta}
    +\frac{(r\xi_{\rm h}b_{\rm r})''b_{\rm r}\xi_{\rm h}^*}{r}H_\varphi^2\cos^2\theta\nonumber\\
&&+\frac{(\xi_{\rm h}b_\theta)'b_{\rm r}\xi_{\rm h}^*}{r}H_\varphi\cos\theta(H_\varphi\sin\theta)'.
\label{FullPoloidal}
\end{eqnarray}

\subsection{Toroidal terms}
\label{sec:tor}

The terms involving the toroidal component are derived:
\begin{eqnarray}
&&2\frac{(r\xi_{\rm r}b_\varphi)'b_\varphi\xi_{\rm h}^*}{r^2}H_{\rm r}H_\theta\sin\theta\cos\theta
    +2\frac{b_\varphi^2|\xi_{\rm h}|^2}{r^2}H_\theta\cos\theta(H_\theta\sin\theta)'\nonumber\\
&&+2m\frac{b_\varphi^2|\xi_{\rm h}|^2}{r^2}H_\theta H_\varphi\cos\theta
    +\frac{(r\xi_{\rm r}b_\varphi)''b_\varphi\xi_{\rm r}^*}{r}H_{\rm r}^2\sin^2\theta\nonumber\\
&&+\frac{(\xi_{\rm h}b_\varphi)'b_\varphi\xi_{\rm r}^*}{r}H_{\rm r}\sin\theta(H_\theta\sin\theta)'
    +\frac{(r\xi_{\rm r}b_\varphi)'b_\varphi\xi_{\rm h}^*}{r^2}H_\theta(H_{\rm r}\sin^2\theta)'\nonumber\\
&&+\frac{b_\varphi^2|\xi_{\rm h}|^2}{r^2}H_\theta[\sin\theta(H_\theta\sin\theta)']'
    -m^2\frac{b_\varphi^2|\xi_{\rm h}|^2}{r^2}H_\theta^2\nonumber\\
&&+\frac{(r\xi_{\rm r}b_\varphi)'(rb_\varphi)'\xi_{\rm r}^*}{r^2}H_{\rm r}^2\sin^2\theta
    +\frac{\xi_{\rm h}b_\varphi(rb_\varphi)'\xi_{\rm r}^*}{r^2}H_{\rm r}\sin\theta(H_\theta\sin\theta)'\nonumber\\
&&+m\frac{\xi_{\rm r}b_\varphi(rb_\varphi)'\xi_{\rm h}^*}{r^2}H_{\rm r}H_\varphi\sin\theta
    -m^2\frac{b_\varphi^2|\xi_{\rm r}|^2}{r^2}H_{\rm r}^2.
\label{FullToroidal}
\end{eqnarray}
\section{JWKB method for ordinary differential equations and rapidly oscillating integrals}
\label{appendix:jwkb}
On the one hand, studying the dynamics of asymptotic low-frequency g and high-frequency p modes implies to solve Schrodinger-like equations of the form
\begin{equation}
\frac{\mathrm{d}^2\Psi\left(x\right)}{\mathrm{d}x^2}+\lambda^2{\hat k}_{x}^{2}\left(x\right)\Psi\left(x\right)=0,
\label{schro4}
\end{equation}
where $\lambda$ is a large parameter and ${\hat k}_{x}$ is a normalized vertical wave number. Applying the JWKB (for Jeffreys-Wentzel-Krammer-Brillouin) method leads for $\lambda\!>\!\!>\!1$ to the following solution \citep[e.g.][]{Erdalyi1956,FromanFroman2005}
\begin{eqnarray}
\Psi\left(x\right)&=&\frac{1}{\sqrt{{\hat k}_{x}\left(x\right)}}
\left[A_{+}\exp\left(i\lambda\int^{x}{\hat k}_{x}\left(x'\right)\,\mathrm{d}x'\right)\right.\nonumber\\
&&\left.+\,A_{-}\exp\left(-i\lambda\int^{x}{\hat k}_{x}\left(x'\right)\,\mathrm{d}x'\right)\right]\,.
\end{eqnarray}
On the other hand, we have to compute integrals of the form
\begin{equation}
I=\int_{a}^{b}g\left(x\right)\exp\left[i\lambda f\left(x\right)\right]{\rm d}x
\end{equation}
to calculate their magnetic splittings. In the cases where $\lambda\!>\!\!>\!1$, they can be approximated by
\begin{equation}
I\rightarrow_{\lambda\rightarrow\infty}\sum_{\vert x_i\vert}g\left(x_i\right)\sqrt{\frac{2\pi}{\vert f"\left(x_i\right)\vert}}\exp\left[i\lambda\left(f\left(x_i\right)-\frac{\pi}{4}{\rm sgn}\left(f"\left(x_i\right)\right)\right)\right],    
\end{equation}
where the $x_i$ are the points such that $f'\left(x_i\right)=0$ \citep[][]{Erdalyi1956}. If $f\left(x\right)=\int^{x}{\hat k}_{x}\left(x'\right)\,\mathrm{d}x'$, the $x_i$ are the turning points such that $\left({\hat k}_{x}\left(x_i\right)\right)^{2}=0$.

\section{Algebra to compute horizontal integrals}

\subsection{Spherical harmonics}

\subsubsection{Definitions}

The spherical harmonics are defined by \citep[e.g.][]{Varshalovischetal1988}:
\begin{equation}
Y_{l}^{m}(\theta,\varphi)=\mathcal{N}_{l}^{m}P_{l}^{|m|}\left(\cos\theta\right)e^{im\varphi},
\label{def}
\end{equation}
where $P_{l}^{|m|}\left(\cos\theta\right)$ is the associated Legendre function, and $\mathcal{N}_{l}^{m}$ the normalization coefficient
\begin{equation}
\mathcal{N}_{l}^{m}=(-1)^{\frac{\left(m+|m|\right)}{2}}\left[\frac{2l+1}{4\pi}\frac{(l-|m|)!}{(l+|m|)!}\right]^{\frac{1}{2}}.
\label{norm}
\end{equation}
They obey the orthogonality relation:
\begin{equation}
\int_{\Omega}\left(Y_{l_{1}}^{m_{1}}\left(\theta,\varphi\right)\right)^{*}Y_{l_{2}}^{m_{2}}\left(\theta,\varphi\right){\rm d}\Omega=\delta_{l_{1},l_{2}}\delta_{m_{1},m_{2}},
\label{ortho}
\end{equation}
where ${\rm d}\Omega=\sin\theta \, {\rm d}\theta \, {\rm d}\varphi$ and where the complex conjugate spherical harmonic is given by:
\begin{equation}\left(Y_{l}^{m}\left(\theta,\varphi\right)\right)^{*}=\left(-1\right)^{m}Y_{l}^{-m}\left(\theta,\varphi\right).
\label{conj}
\end{equation}

\subsubsection{Linear recurrence relations}
We introduce useful linear recurrence relations that should be used to compute analytically the $C_{l,m}$ and $D_{l,m}$ coefficients. Using again \cite{Varshalovischetal1988}, we have on the one hand 
\begin{equation}
\cos\theta\,Y_l^m=\alpha_{l,m}Y_{l+1}^{m}+\beta_{l,m}Y_{l-1}^{m},
\end{equation}
where
\begin{equation}
\alpha_{l,m}=\left[\frac{\left(l+1+m\right)\left(l+1-m\right)}{\left(2l+1\right)\left(2l+3\right)}\right]^{1/2}   
\end{equation}
and
\begin{equation}
\beta_{l,m}=
\left[\frac{\left(l+m\right)\left(l-m\right)}{\left(2l+1\right)\left(2l-1\right)}\right]^{1/2}.
\end{equation}
On the other hand, we get
\begin{equation}
\partial_{\theta}Y_l^m=A_{l,m}Y_{l}^{m+1}e^{-i\varphi}-B_{l,m}Y_{l}^{m-1}e^{i\varphi}
\label{R1}
\end{equation}
\begin{equation}
-m\frac{\cos\theta}{\sin\theta}Y_l^m=A_{l,m}Y_{l}^{m+1}e^{-i\varphi}+B_{l,m}Y_{l}^{m-1}e^{i\varphi}    
\label{R2}
\end{equation}
where
\begin{equation}
A_{l,m}=\frac{1}{2}\sqrt{\left[l\left(l+1\right)-m\left(m+1\right)\right]}  
\end{equation}
and
\begin{equation}
B_{l,m}=\frac{1}{2}\sqrt{\left[l\left(l+1\right)-m\left(m-1\right)\right]}.    
\end{equation}

\subsubsection{Products of spherical functions and specific integrals}
Integrals involving products of spherical harmonics should also be computed to calculate $C_{l,m}$ and $D_{l,m}$. Using the normalization and the orthogonality of spherical harmonics (Eqs. \ref{norm}-\ref{ortho}) and their complex conjugate (Eq. \ref{conj}), we can write:
\begin{eqnarray}
\lefteqn{Y_{l_{1}}^{m_{1}}\left(\theta,\varphi\right)Y_{l_{2}}^{m_{2}}\left(\theta,\varphi\right)}\nonumber\\
&=&\sum_{l=|l_{1}-l_{2}|}^{l_{1}+l_{2}}c_{l_{1},m_{1},l_{2},m_{2}}^{l}Y_{l}^{m_{1}+m_{2}}\left(\theta,\varphi\right)\nonumber\\
&=&(-1)^{\left(m_{1}+m_{2}\right)}\sum_{l=|l_{1}-l_{2}|}^{l_{1}+l_{2}}\mathcal{I}_{l_{1},l_{2},l}^{m_{1},m_{2},-\left(m_{1}+m_{2}\right)}Y_{l}^{m_{1}+m_{2}}\left(\theta,\varphi\right),
\end{eqnarray}
where we define the integral $\mathcal{I}_{l_{1},l_{2},l}^{m_{1},m_{2},m}$ like in \cite{Varshalovischetal1988}:
\begin{eqnarray}
\lefteqn{\mathcal{I}_{l_{1},l_{2},l}^{m_{1},m_{2},m}=\int_{\Omega}Y_{l_{1}}^{m_{1}}\left(\theta,\varphi\right)Y_{l_{2}}^{m_{2}}\left(\theta,\varphi\right)Y_{l}^{m}\left(\theta,\varphi\right)d\Omega}\nonumber\\
&=&\sqrt{\frac{(2l_{1}+1)(2l_{2}+1)(2l+1)}{4\pi}}\left(\begin{array}{ccc}
l_{1} & l_{2} & l \\
m_{1} & m_{2} & m
\end{array}\right)
\left(\begin{array}{ccc}
l_{1} & l_{2} & l\\
0 & 0 & 0
\end{array}\right)\nonumber\\
\label{Ylmcouplings}
\end{eqnarray}
with the 3j-Wigner coefficients $\left(\cdot\!\cdot\!\cdot\right)$ that are related to the classical Clebsch-Gordan coefficients by: 
\begin{equation}
\left(\begin{array}{ccc}
l_{1} & l_{2} & l \\
m_{1} & m_{2} & m 
\end{array}\right)=\frac{(-1)^{l_{1}-l_{2}-m}}{\sqrt{2l+1}}C_{l_{1},m_{1},l_{2},m_{2}}^{l,-m}.
\end{equation}
Then, using the initial definition of spherical harmonics (cf. Eqs. \ref{def}-\ref{norm}), we deduce the expansion for the product of two associated Legendre functions: 
\begin{equation}
P_{l_{1}}^{m_{1}}\left(\cos\theta\right)P_{l_{2}}^{m_{2}}\left(\cos\theta\right)=\sum_{l=|l_{1}-l_{2}|}^{l_{1}+l_{2}}d_{l_{1},m_{1},l_{2},m_{2}}^{l}P_{l}^{m_{1}+m_{2}}\left(\cos\theta\right),
\end{equation}
where
\begin{eqnarray}
\lefteqn{d_{l_{1},m_{1},l_{2},m_{2}}^{l}=(-1)^{\left(m_{1}+m_{2}\right)}\left(2l+1\right)}\nonumber\\
&&\times\sqrt{\frac{\left(l_{1}+m_{1}\right)!\left(l_{2}+m_{2}\right)!\left(l-\left(m_{1}+m_{2}\right)\right)!}{\left(l_{1}-m_{1}\right)!\left(l_{2}-m_{2}\right)!\left(l+\left(m_{1}+m_{2}\right)\right)!}}\nonumber\\
&&\times\left(\begin{array}{ccc}
l_{1} & l_{2} & l \\
m_{1} & m_{2} & -\left(m_{1}+m_{2}\right)
\end{array}\right)
\left(\begin{array}{ccc}
l_{1} & l_{2} & l\\
0 & 0 & 0
\end{array}\right).
\label{couplingPlm}
\end{eqnarray}
The last relevant integral is
\begin{eqnarray}
\lefteqn{\int_{-1}^{1}P_{l}^{m}\left(\mu\right){\rm d}\mu=\frac{\displaystyle{\left(-1\right)^{l}+\left(-1\right)^{m}}}{\displaystyle{\left(\frac{l-m}{2}\right)!\,\Gamma\left(\frac{l+3}{2}\right)}}}\nonumber\\
&\times&2^{m-2}m\Gamma\left(\frac{l}{2}\right)\Gamma\left(\frac{1}{2}\left(l+m+1\right)\right),
\label{IntegralPlm}
\end{eqnarray}
where $\Gamma$ are the usual Gamma functions \citep{AbramowitzStegun1972}.

\subsection{Application to $C_{l,m}$ coefficients}
It is possible to provide a fully analytical expression of $C_{l,m}$ by using Eqs. (\ref{R1}, \ref{R2}, \ref{couplingPlm}, \ref{IntegralPlm}). However, it is very cumbersome and the only interesting tractable expression is obtained in the case when $l=m$ where
\begin{eqnarray}
C_{l,l}&=&\frac{1}{l\left(l+1\right)}\left[A_{l,l}^{2}\left[\left(\alpha_{l,l+1}\right)^2+\left(\beta_{l,l+1}\right)^2+1\right]\right.\nonumber\\
&&+\left.B_{l,l}^{2}\left[\left(\alpha_{l,l-1}\right)^2+\left(\beta_{l,l-1}\right)^2+1\right]\right]\nonumber\\
&=&\frac{3+l}{3+5l+2l^2}.
\end{eqnarray}

The values of $C_{l,m}$ for the dipolar ($l=1$) and the quadrupolar ($l=2$) modes are
\begin{equation}
C_{1,0}=\frac{1}{5},\quad C_{1,1}=C_{1,-1}=\frac{2}{5},
\end{equation}
\begin{equation}
C_{2,0}=\frac{9}{21},\quad C_{2,1}=C_{2,-1}=\frac{8}{21}, \quad C_{2,2}=C_{2,-2}=\frac{5}{21}.
\end{equation}

\subsection{Application to $D_{l,m}$ coefficients}
\label{Dcompute}

We have to compute
\begin{equation}
D_{l,m}=\int_{\Omega}\sin^2\theta\vert Y_l^m\vert^2{\rm d}\Omega.
\end{equation}
Using Eq. (\ref{conj}) and the identity
\begin{equation}
\sin^2\theta=\frac{2}{3}\left(1-\frac{Y_{2}^{0}}{{\mathcal N}_{2}^{0}}\right),
\end{equation}
where ${\mathcal N}_{2}^{0}=1/2\sqrt{5/\pi}$ is obtained thanks to Eq. (\ref{norm}), we get
\begin{equation}
D_{l,m}=\frac{2}{3}\left[1-\frac{\left(-1\right)^{m}}{{\mathcal N}_{2}^{0}}{\mathcal I}_{2,l,l}^{0,m,-m}\right],
\end{equation}
where ${\mathcal I}$ has been defined in Eq. (\ref{Ylmcouplings}).

The values of $D_{l,m}$ for the dipolar ($l=1$) and the quadrupolar ($l=2$) modes are
\begin{equation}
D_{1,0}=\frac{2}{5},\quad D_{1,1}=D_{1,-1}=\frac{4}{5},
\end{equation}
\begin{equation}
D_{2,0}=\frac{10}{21},\quad D_{2,1}=D_{2,-1}=\frac{4}{7}, \quad D_{2,2}=D_{2,-2}=\frac{6}{7}.
\end{equation}
\end{appendix}
\end{document}